\documentclass[twocolumn,aps,prb]{revtex4}
\usepackage{graphicx,graphics,color,epsfig}
\usepackage{amsmath}
\usepackage{amssymb}
\usepackage[small]{caption}

\begin{document}
\title{
Nonequilibrium dynamic-correlation-length scaling method
}
\author{Tota Nakamura}
\affiliation{%
Faculty of Engineering,
Shibaura Institute of Technology,  
307 Fukasaku, Minuma, Saitama 337-8570, Japan}

\date{\today}

\begin{abstract}
The finite-size scaling method in the equilibrium Monte Carlo(MC) 
simulations and
the finite-time scaling method in the nonequilibrium-relaxation simulations
are compromised.
MC time data of various physical quantities are scaled by the
MC time data of the dynamic correlation length, which corresponds to changing 
the system size in the finite-size scaling method.
This scaling method is tested in the three-dimensional ferromagnetic Ising  
spin model and in the three dimensional $\pm J$ Ising spin-glass model.
The transition temperature and the critical exponents, $\eta$ and $\nu$, 
are obtained by the nonequilibrium relaxation data of the susceptibility and
the dynamic correlation length apart from the dynamic exponent.
We also comment on the definition of the dynamic correlation length in the
nonequilibrium relaxation process.
The Ornstein-Zernike formula is not always appropriate.
\end{abstract}
\maketitle

\section{Introduction}

A Monte Carlo(MC) simulation 
and the finite-size-scaling analysis
are very important tools in the study of phase transitions.\cite{binder}
The applications serve as a strong bridge between the experimental and the
theoretical physics.
Using this method,
we may estimate various physical parameters, predict unknown properties, and
propose new experiments on real materials.
One of successful applications may be the study on the phase transition.
We perform the finite-size-scaling analysis on the numerical data obtained by
the MC simulations.
Then, the critical temperature and the critical exponents are estimated.

The nonequilibrium relaxation method\cite{Stauffer,Ito,nerreview,ItoOz1,OzIto2}
is an alternate version of the MC simulation that
studies the phase transitions.
The method directly deals with the MC 
relaxation functions of physical quantities.
In this scheme, time and size are related by the dynamic scaling relation:
$\tau\sim \xi^z$, which connects the correlation time $\tau$ and
the correlation length $\xi$ via the dynamic exponent $z$.
In the standard equilibrium simulations,
we take the infinite-time (equilibrium) limit first. 
Then, infinite-size limit
is taken by the finite-size-scaling analysis.
This procedure is reversed in the nonequilibrium relaxation method.
We take the infinite-size limit first
by preparing a very large system and stopping the simulation
before the finite-size effects appear.
Then, infinite-time (equilibrium) limit is taken by the
finite-time-scaling analysis.
The critical temperature and the critical exponents are estimated in the
same manner as the standard finite-size-scaling analysis.

An advantage of the nonequilibrium relaxation method
is a computational efficiency.
We use the nonequilibrium MC data that are usually discarded in the 
equilibrium simulations.
It is very important particularly when the computational resources at hand
are limited.

One shortcoming in the nonequilibrium-relaxation method is an ambiguity of the
dynamic exponent, $z$.
We obtain the transition temperature and the critical exponent, $\gamma$, by
the finite-time-scaling analysis on the 
susceptibility.\cite{OzIto2,shirahata1,nakamura,nakamura2,%
shirahata2,yamamoto1,nakamura4}
However,
the exponent $\nu$ is not solely obtained but it appears as a form $z\nu$.
We need to estimate $z$ independently in order to obtain $\nu$.
The dynamic exponent takes a large value in the frustrated and/or random
systems.
It sometimes depends on the temperature in spin-glass models.\cite{katzgraber}
The numerical estimate of $z$ is a tough task in these simulations.

In this paper, we introduce another scaling method based on the nonequilibrium
relaxation scheme.
The method is free from the estimate of $z$.
We focus on the nonequilibrium relaxation function of the correlation length,
which is called the dynamic correlation length.
It serves as the system size in the finite-size-scaling analysis.
The nonequilibrium relaxation functions of various physical quantities are 
scaled by the dynamic correlation length.
We obtain the transition temperature and the critical exponents by this
scaling analysis in the same manner as the finite-size scaling.
We just replace the system size $L$ of the finite-size scaling analysis
with the dynamic correlation length $\xi(t)$, where $t$ is the MC time step.
We also propose a definition of the dynamic correlation length.
The Ornstein-Zernike formula\cite{ornstein} is often used to define this
value.
We show that this definition is not always appropriate particularly
in the nonequilibrium relaxation regime.

Section \ref{sec:defxi} explains models and 
a definition of the dynamic correlation length.
Section \ref{sec:scaling} explains the scaling procedure. 
Numerical results are presented in Sec. \ref{sec:results}.
The ferromagnetic Ising model and the $\pm J$ Ising spin-glass model are
taken as examples.
Section \ref{sec:discussion} is devoted to the summary and discussions.

\section{Model and Definition of the dynamic correlation length}
\label{sec:defxi}

Let us consider the following spin model to demonstrate the present scaling method.
\[
{\cal H} = -\sum_{\langle i, j \rangle} J_{ij} S_i S_j,
\label{eq:hamil}
\]
where $\langle i, j \rangle$ denotes the nearest-neighbor pairs in the
cubic lattice, $J_{ij}$ denotes the exchange interaction, and 
$S_i$ denote the Ising spin. 
We consider the uniform ferromagnetic model ($J_{ij}=1$) and the
spin-glass model ($J_{ij}=\pm J$ with an equal probability) in this paper.

The correlation length of an ordered domain in spin models is 
usually
defined by the correlation function 
$C(r)=\langle S_i S_{i+r} \rangle$ as\cite{ornstein}
\[
C(r)= \langle S_i S_{i+r} \rangle \propto \frac{\exp[-r/\xi]}{r}
\]
in the equilibrium state near the transition temperature.
In the Monte Carlo simulations, it is easier to estimate the correlation length
by the Fourier transform of the susceptibility, 
$\chi({\mbox{\boldmath $k$}})$, as\cite{cooper82}
\begin{equation}
\xi = \frac{1}{k_{\rm min}}\sqrt{
\frac{{\chi}(0)}
     {{\chi}({\mbox{\boldmath $k$}_{\rm min}})}-1
}, ~~~
\mbox{\boldmath $k$}_{\rm min}
=\frac{2\pi}{L}\left(\begin{matrix}1\cr 0\cr 0\end{matrix}\right).
\label{eq:1}
\end  {equation}
Here, $\mbox{\boldmath $k$}_{\rm min}$ denotes the smallest wave vector along
one direction in a finite-size lattice.
The expression is exact in the limit of $\mbox{\boldmath $k$}_{\rm min}\to 0$.

The $1/r$-correction form in the spin correlation function is only guaranteed 
by the mean-field theory
in the equilibrium state.\cite{ornstein}
It is not trivial that we may use Eq.~(\ref{eq:1}) in the nonequilibrium
relaxation process.
The system size $L$ is much larger than the correlation length in the 
nonequilibrium process.
In such case, the spin correlation function may decay in a simple exponential 
form without the $1/r$-correction.
In order to check the $r$-dependence, we plot the correlation functions, 
$C(r)$ and $rC(r)$, against $r$ in the nonequilibrium regime ($t=501$) and in the
equilibrium regime ($t=3981$).
The system is the uniform ferromagnetic model with $J_{ij}=1$, 
the system size $L=299$, and the temperature $T=4.515$, which is above
the transition temperature, $T_{\rm c}=4.511525$.\cite{3dising}
\begin{figure}
  \resizebox{0.45\textwidth}{!}{\includegraphics{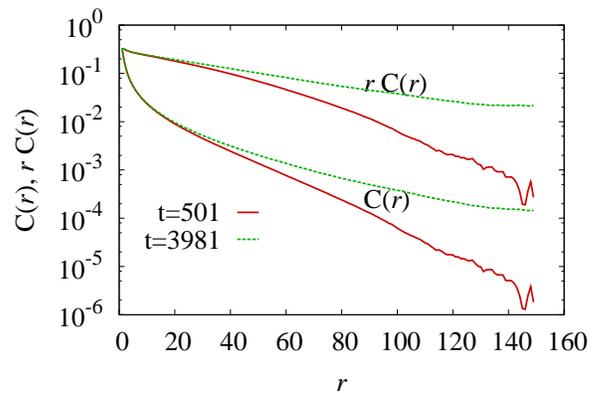}}
  \caption{
(Color online)
$r$ dependence of the spin correlation function in the nonequilibrium 
regime ($t=501$) and in the equilibrium regime ($t=3981$).
The system is the uniform ferromagnetic model in three dimension.
The linear system size $L=299$, and the temperature $T=4.515$.
The lower two lines are plots of $C(r)$, and the upper two lines are
those of $rC(r)$.
}
\label{fig:corr}
\end  {figure}
Figure \ref{fig:corr} shows the result.
The linearity of $rC(r)$ is better than that of $C(r)$ when $t=3981$.
The $1/r$-correction is necessary in estimating the correlation length
in the equilibrium state.
On the other hand,
the linearity is poor and convex upward when $t=501$.
The linearity of $C(r)$ seems to be better.
If the correlation function exhibits the simple exponential decay as
\[
C(r)= \langle S_i S_{i+r} \rangle \propto {\exp[-r/\xi]},
\]
the correlation length should be estimated by
\begin{equation}
\xi = \frac{1}{k_{\rm min}}\sqrt{
\sqrt{
\frac{{\chi}(0)}
     {{\chi}({\mbox{\boldmath $k$}_{\rm min}})}
}
-1
}.
\label{eq:2}
\end  {equation}
It is difficult to judge a use of Eq.~(\ref{eq:1}) or Eq.~(\ref{eq:2})
just by the linearity of $C(r)$ or $rC(r)$.
The difference is small as shown in Fig.~\ref{fig:corr}.
Therefore,
we compare the time dependences of the dynamic correlation length
estimated by both definitions.
We judge two definitions by the scaling behavior.

\begin{figure}
  \resizebox{0.45\textwidth}{!}{\includegraphics{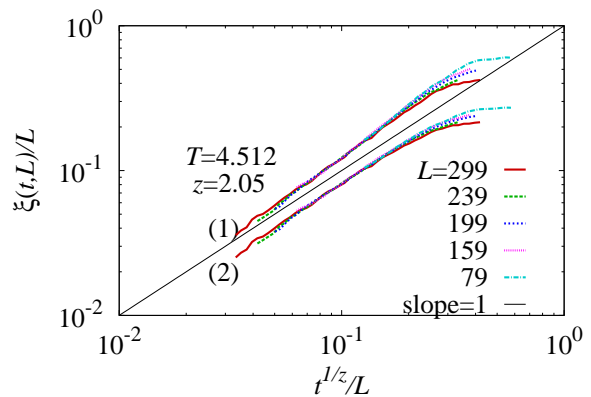}}
  \caption{
(Color online)
The dynamic correlation length $\xi(t, L)$ over $L$ is plotted against
$t^{1/z}$ over $L$.
Data labelled by (1) are estimated by Eq.~(1), and 
those labelled by (2) are estimated by Eq.~(2). 
The system is the uniform ferromagnetic Ising model in three dimensions.
The temperature $T=4.512$ is very close to the transition temperature.
The estimate of the dynamic exponent $z=2.05$ is taken from the 
Ref.~\cite{3dising}.
}
\label{fig:isingxi}
\end  {figure}
\begin{figure}
  \resizebox{0.45\textwidth}{!}{\includegraphics{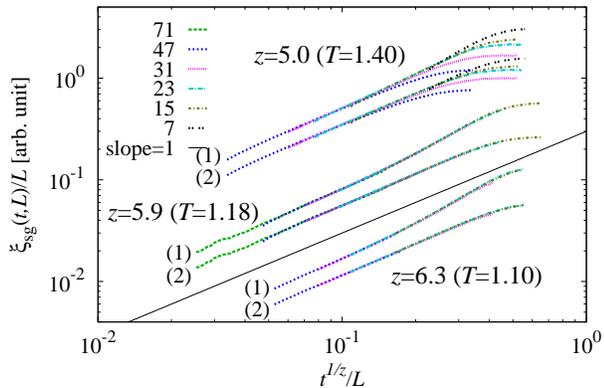}}
  \caption{
(Color online)
The dynamic correlation length $\xi(t, L)$ over $L$ is plotted against
$t^{1/z}$ over $L$ in the $\pm J$ Ising spin-glass model in three dimensions.
Data labelled by (1) are estimated by Eq.~(1), and 
those labelled by (2) are estimated by Eq.~(2). 
Three temperatures, $T=1.4$, 1.18, and 1.10, are above, near, and below the
transition temperature, respectively.
We estimated the dynamic exponent so that the dynamic correlation length
are well-scaled at each temperature.
}
\label{fig:sgxi}
\end  {figure}

Figures \ref{fig:isingxi} and \ref{fig:sgxi} compare the estimates of 
$\xi(t)$ using both definitions, Eqs. (\ref{eq:1}) and (\ref{eq:2}).
The dynamic correlation length of the size-$L$ system, $\xi(t, L)$
divided by the linear scale $L$ is plotted
against $t^{1/z}/L$.
We expect that $\xi(t, L) \propto t^{1/z}$ from the dynamic scaling relation.
The correlation length divided by $L$ becomes scale invariant at the 
transition temperature.
Therefore, the present plot should ride on the single line with its slope unity,
if we choose a correct value of $z$ that satisfies the dynamic scaling relation.

In the uniform ferromagnetic Ising model (Fig. \ref{fig:isingxi}),
the dynamic correlation length estimated by Eq.~(\ref{eq:2})
yields a good scaling and the slope unity.
The dynamic exponent $z=2.05$ is taken from the estimate given by
Ito {\it et al.}\cite{3dising}.
The finite-size effects appear as the system approaches the equilibrium state.

The dynamic correlation length given by Eq.~(\ref{eq:1}) 
is well-scaled but the slope does not agree with unity.
If we choose $z$ so that it yields the slope unity, the relaxation
data of different sizes do not ride on the same line.
The slope suggests that $\xi(t,L)/ L\propto (t^{1/z}/L)^{1.13}$, but
the scaling fails if we take $z=2.05/1.13=1.81$ for $t^{1/z}/L$.
Therefore, the dynamic scaling relation contradicts itself
even in the ferromagnetic model.
It is an evidence that a use of
the Ornstein-Zernike expression, Eq.~(\ref{eq:1}), is not appropriate
in the nonequilibrium relaxation scheme.

Figure \ref{fig:sgxi} shows the same $\xi(t, L)/L$ scaling 
in the $\pm J$ Ising spin-glass model in three dimensions.
Relaxation data at three temperatures are plotted in the same figure
by shifting arbitrary in the vertical direction.
The temperatures are above ($T=1.40$), near ($T=1.18$), and below ($T=1.10$)
the transition temperature.
The dynamic exponent at each temperature is estimated so that the relaxation 
data fall onto the same line with the slope unity.
The linearity is better if we use Eq.~(\ref{eq:2}).
The estimate of $\xi(t, L)$ using Eq.~(\ref{eq:1}) bends upward
as the system approaches the equilibrium limit, while the slope approaches
unity in the nonequilibrium limit ($t\to 0$).
This bending behavior may mislead us to underestimate the value of $z$.
It is also noted that the value of $z$
depends on the temperature as $zT\simeq 7$.\cite{katzgraber}
The finite-size effects appear when the temperature is above the transition
temperature ($T=1.40$).
The size effects are weak in the spin-glass phase ($T=1.18$ and 1.10).

It is shown that the dynamic correlation length in the nonequilibrium 
relaxation process should be estimated by using Eq.~(\ref{eq:2}).
The dynamic scaling relation is satisfied in the nonequilibrium process if we
take this definition.
We perform the dynamic correlation-length scaling analysis using
Eq.~(\ref{eq:2}) in this paper.

\section{Scaling procedure}
\label{sec:scaling}

We explain our scaling procedures step by step in this section.
The first two steps below are the standard 
procedure of the nonequilibrium relaxation method.
The relaxation functions of the physical quantity and the correlation length
are obtained.

\begin{enumerate}
\item
We prepare a system with a linear scale $L$, and perform a MC simulation at
the temperature $T$.
We start the simulation from the paramagnetic state.
The correlation length is zero in the initial state.
A physical quantity, $A$, is measured at each MC step, $t$, and is stored 
in memory as $A(t, L, T)$.
Changing a random number sequence, we perform independent simulations and 
take an average of $A(t, L, T)$ over these runs.

\item
We change the system size and perform simulations.
Relaxation functions $A(t, L, T)$ of different sizes are compared
to check the finite-size effects.
Only data that are free from the finite-size effects 
are used in the scaling analysis.
Examples are Figs. 4 and 7 in Sec.~\ref{sec:results}.
The data free from the size effects are denoted by $A(t, T)$.
We obtain a relaxation function of the dynamic correlation length,
$\xi(t, T)$, as explained in Sec.~\ref{sec:defxi}.

\end{enumerate}

We will perform three scaling analyses below.
The first scaling analysis determines the dynamic exponent, $z$.
Then, the second one determines the anomalous exponent $\eta$, and the
last scaling analysis determines the critical temperature, $T_{\rm c}$, and
the exponent $\nu$.
It is noted that the first scaling analysis is not necessary and is 
independent from the other two scaling analyses.
It should be done only when we need a value of $z$.
This is a clear difference from (and is probably an advantage over) the standard
nonequilibrium relaxation method with the finite-time-scaling analysis.

\begin{enumerate}
\setcounter{enumi}{2}
\item
The relaxation function of the dynamic correlation length should 
satisfy the dynamic scaling relation:
$\xi(t, T)\sim t^{1/z}$.
At the temperature near the critical temperature,
We plot $\xi(t,L)/L$ versus $t^{1/z}/L$ so that all the scaled data fall on
a single scaling function.
This scaling is possible because $\xi/L$ is scale-invariant at the critical
temperature.
Only a value of $z$ is a control parameter.
Examples are Figs. 2 and 3.

\end{enumerate}

Let us take the magnetic susceptibility, $\chi$, as a physical quantity for
simplicity.

\begin{enumerate}
\setcounter{enumi}{3}
\item
Using the relaxation function of $\chi(t, T)$ and $\xi(t, T)$, we perform
the scaling analysis.
Since the simulation starts from the paramagnetic state, both $\chi$ and $\xi$
are expected to increase algebraically in time as 
$\chi(t, T) \sim \xi(t, T)^{2-\eta}$ when $T$ is near $T_{\rm c}$.
Since $\chi \propto \xi^{2-\eta}$ in the critical region,
we expect that the nonequilibrium relaxation functions
of $\chi(t, L)$ and $\xi(t, L)$ depend on $L$ as
\begin{equation}
\frac{\chi(t, L)}{L^{2-\eta}} \propto \left(\frac{\xi(t, L)}{L}\right)^{2-\eta}.
\label{eq:chi-xi}
\end{equation}
We plot $\chi(t, L)/L^{2-\eta}$ against $\xi(t, L)/L$ in a log-log scale
and determine $\eta$ so that all the relaxation functions falls on the 
same line with a slope $2-\eta$.
Examples are Figs. 5 and 8 in Sec.~\ref{sec:results}.
It is a one-parameter scaling analysis.
The determination of $\eta$ is easily done first by this scaling plot.
Here, it is not necessary to know the critical temperature precisely.
We only need the rough estimate.
Since we work with the nonequilibrium relaxation method, the critical relaxation
is observed near the critical temperature in the nonequilibrium process.
It is sufficient to perform this scaling analysis at one temperature
near $T_{\rm c}$.

\item
The critical temperature, $T_{\rm c}$, and the critical exponent, $\nu$, are
determined in the last scaling analysis.
The equilibrium value of the correlation length diverges as
$\xi\sim |T-T_{\rm c}|^{-\nu}$.
At the same time the susceptibility $\chi(t, T)$ should be scaled by
$\xi(t, T)^{2-\eta}$ with $\eta$ estimated above.
Therefore, we plot the relaxation functions
$\chi(t, T)/\xi(t, T)^{2-\eta}$ versus $\xi(t, T)/|T-T_{\rm c}|^{-\nu}$ for
various temperatures and determine $T_{\rm c}$ and $\nu$ so that all the 
data ride on a single scaling function.
Examples are Figs. 6 and 9. in Sec.~\ref{sec:results}.

\end  {enumerate}

It is noted that the finite size $L$ in the finite-size-scaling analysis
is replaced by the dynamic correlation length
$\xi(t, T)$ in the present analysis.
We consider that the present scaling analysis is a natural extension 
from the finite-size-scaling analysis
when we work within the nonequilibrium relaxation scheme.

\section{Numerical results}
\label{sec:results}

\subsection{Ferromagnetic model}

\begin{figure}
  \resizebox{0.45\textwidth}{!}{\includegraphics{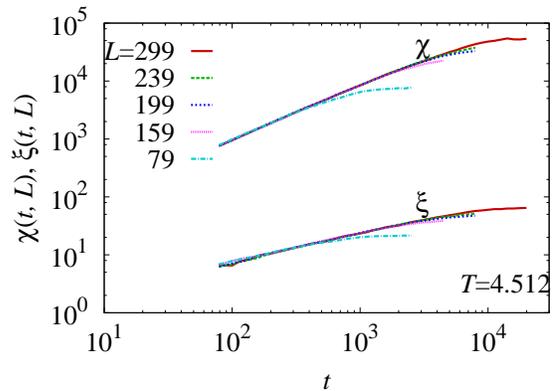}}
  \caption{
(Color online)
Size dependences of 
raw relaxation data of the dynamic correlation length, $\xi(t, L)$, and
the magnetic susceptibility, $\chi(t, L)$ in the ferromagnetic model.
The temperature, $T=4.512$, is a little above the transition temperature.
}
\label{fig:chixi-ferro-nama}
\end  {figure}

\begin{figure}
  \resizebox{0.4\textwidth}{!}{\includegraphics{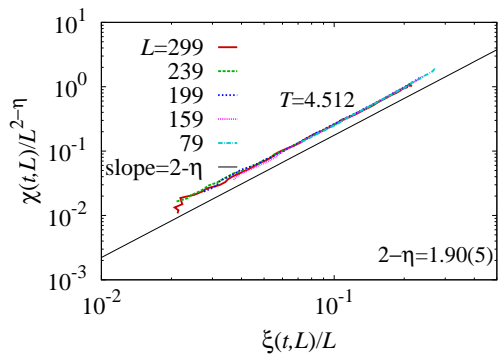}}
  \caption{
(Color online)
The $\chi-\xi$ scaling plot of Eq.~(\ref{eq:chi-xi}) 
in the ferromagnetic model.
The raw relaxation data of Fig.~\ref{fig:chixi-ferro-nama} are scaled.
}
\label{fig:chixi-ferro}
\end  {figure}

We present the numerical results in the ferromagnetic Ising model in 
three dimension.
Figure \ref{fig:chixi-ferro-nama} shows the raw relaxation data of the
dynamic correlation length and the magnetic susceptibility.
The dynamic correlation length is estimated using Eq.~(\ref{eq:2}).
We plot in Fig.~\ref{fig:chixi-ferro}
the scaling result that determines $\eta$.
The relaxation data are scaled with $2-\eta=1.90\pm 0.05$, where 
the logarithmic slope of the scaled function agrees with $2-\eta$.

\begin{figure}
  \resizebox{0.45\textwidth}{!}{\includegraphics{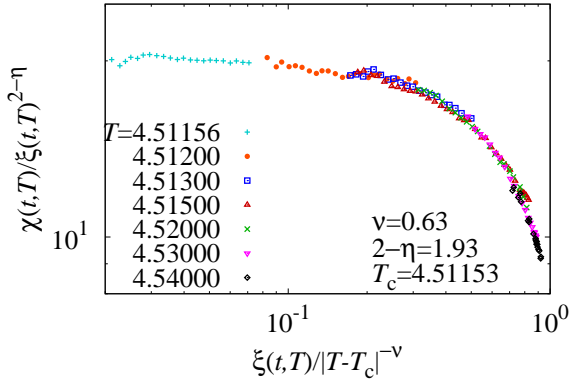}}
  \caption{
(Color online)
The dynamic correlation-length scaling plot of the magnetic susceptibility 
in the ferromagnetic model.
}
\label{fig:xichi-ferro}
\end  {figure}

The dynamic correlation-length scaling of the susceptibility is shown in 
Fig.~\ref{fig:xichi-ferro}.
Here, only the nonequilibrium-relaxation data that are free from the finite-size
effects are plotted.
For a estimated value of $(2-\eta)$ between $1.85$ and $1.95$, we search for
values of $T_c$ and $\nu$ so that the scaled data yield the best scaling behavior.
The obtained values are listed in Fig.~\ref{fig:xichi-ferro} and
are consistent with the previous estimates.\cite{3dising}

\subsection{Spin-glass model}

In this subsection, we present the numerical results of the $\pm J$ 
Ising spin-glass model in three dimensions.
The spin-glass correlation length is estimated from the Fourier transform
of the spin-glass susceptibility, $\chi_{\rm sg}$. 
It is defined by
\[
\chi_{\rm sg}=\frac{1}{N}\left[
\sum_{i,j}\langle S_i S_j \rangle ^2
\right]_{\rm c},
\]
where the bracket $[\cdots ]_{\rm c}$ denotes the configurational average,
and the bracket $\langle \cdots \rangle$ denotes the thermal average.
The thermal average is estimated by the average over different real replicas:
\[
\langle S_i S_j \rangle=\frac{1}{m}\sum_{A=1}^m S_i^{(A)} S_j^{(A)}.
\]
A replica number is denoted by $m$, and the superscript $A$ denotes a
replica index.
We prepare $m$ real replicas for each random bond realization with
different initial spin configurations.
Spin states of each replica are updated using different random number 
sequences.
The thermal average is taken only by this replica average in our
nonequilibrium relaxation scheme.
A replica number controls an accuracy of the thermal average.
It is set to 256 in this paper.
The dynamic spin-glass correlation length in this paper is estimated by
Eq.~(\ref{eq:2}), where $\chi$ is replaced by 
$\chi_{\rm sg}$.
It is usually defined by
Eq.~(\ref{eq:1}).\cite{palassini,ballesteros}

\begin{figure}
  \resizebox{0.45\textwidth}{!}{\includegraphics{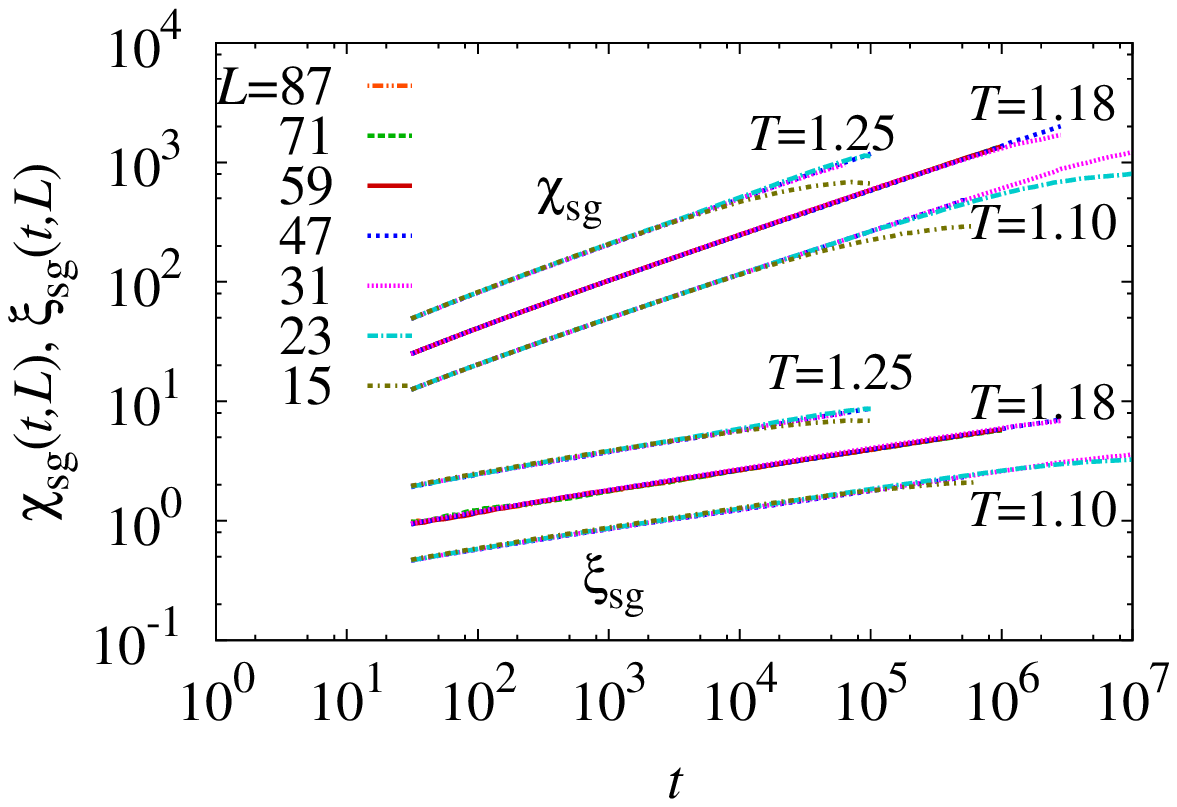}}
  \caption{
(Color online)
Size dependences of 
raw relaxation data of the dynamic correlation length, $\xi_{\rm sg}(t, L)$, and
the spin-glass susceptibility, $\chi_{\rm sg}(t, L)$ in the spin-glass model.
Three temperatures are above, near, and below the transition temperature,
respectively.
Data of $T=1.10$ are divided by 2, and those of $T=1.25$ are multiplied
by 2 in order to separate from data of $T=1.18$.
}
\label{fig:chixi-nama}
\end  {figure}

\begin{figure}
  \resizebox{0.45\textwidth}{!}{\includegraphics{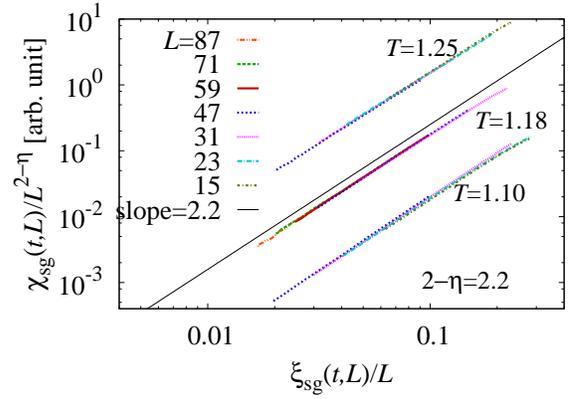}}
  \caption{
(Color online)
The $\chi_{\rm sg}$-$\xi_{\rm sg}$ scaling of Eq.~(\ref{eq:chi-xi}) in the
spin-glass model.
The raw relaxation data of Fig.~\ref{fig:chixi-nama} are scaled.
A straight line is a guide for eyes with a slope $2-\eta=2.2$.
}
\label{fig:chixi}
\end  {figure}

\begin{figure}
  \resizebox{0.45\textwidth}{!}{\includegraphics{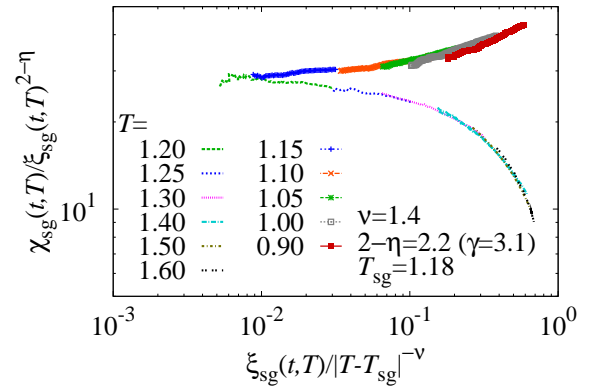}}
  \caption{
(Color online)
The dynamic correlation-length scaling plot of the spin-glass susceptibility.
}
\label{fig:xichi}
\end  {figure}

Figure \ref{fig:chixi-nama} shows the raw relaxation functions of the
spin-glass susceptibility and the dynamic spin-glass correlation length.
Figure \ref{fig:chixi} shows the scaling plot determining a value of $\eta$.
The straight-line scaling is possible near and 
above the transition temperature.
The scaling poor in the low-temperature  phase.
Using the estimated value of $2-\eta$ we plot the dynamic correlation-length
scaling results in Fig.~\ref{fig:xichi}.
The scaling is good when the temperature is above the estimated 
spin-glass transition temperature, $T_{\rm sg}=1.18$.
The scaling behavior becomes poor below $T_{\rm sg}$.
The estimated critical exponents may be invalid in the low-temperature
phase.

The previous estimates\cite{ballesteros,Bhatt,Ogielski,KawashimaY,%
maricampbell}
for $T_{\rm sg}$ lie between 1.1 and 1.2.
They are roughly categorized into two groups.
One\cite{ballesteros,KawashimaY} gives $T_{\rm sg}$ close to 1.1 
and $\nu$ close to 2.
The other\cite{Bhatt,Ogielski,maricampbell}
gives $T_{\rm sg}$ close to 1.2 and $\nu$ close to 1.3.
The present result is consistent with the latter group.
The latter group mostly takes the dynamic approach to the phase transition.
Recently, Hukushima and Campbell\cite{hukucampcorrection}
discussed that this discrepancy can be
understood by the strong corrections to scaling.

\begin{figure}
  \resizebox{0.45\textwidth}{!}{\includegraphics{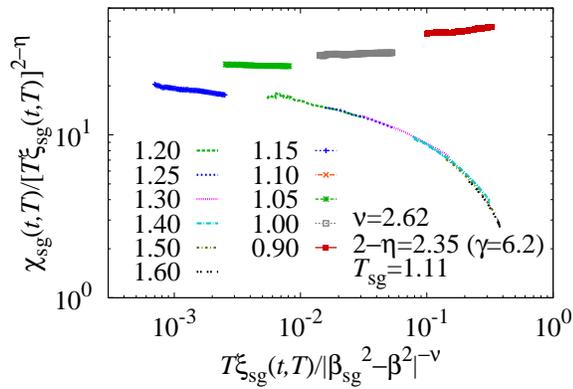}}
  \caption{
(Color online)
The dynamic correlation-length scaling plot of the spin-glass susceptibility
using $|\beta_{\rm c}^2-\beta^2|$ as the scaling variable.
}
\label{fig:betascale}
\end  {figure}

Campbell {\it et~al.} \cite{betascale}
proposed the $\beta$-scaling analysis, which 
uses $(\beta^2-\beta_{\rm c}^2)$ as the scaling variable.
They estimated the transition temperature and the critical exponents as
$T_{\rm c}=1.11$, $\nu=2.72$, and $2-\eta=2.40(4)$.
We also try this $\beta$-scaling analysis in Fig. \ref{fig:betascale}.
Our estimates are
$T_{\rm c}=1.11$, $\nu=2.62$, and $2-\eta=2.35$, which are consistent
with their estimates but disagree with our present estimates using 
$(T-T_{\rm c})$.
This discrepancy suggests 
that the present numerical simulation is not 
sufficient to extract the true critical phenomena both in a size scale
and a time scale in the spin-glass model.\cite{hukucampcorrection}
We checked that the $\beta$-scaling is possible
using the same transition temperature and critical exponents
in the ferromagnetic model.
(Figures are not shown.)
The present size and time scales are sufficient in the ferromagnetic model.

\section{Summary and Discussion}
\label{sec:discussion}

The dynamic correlation-length-scaling method is introduced.
The basic idea of this method is that we investigate the phase transition
through the correlation length.
It is found that the scaling relations among physical quantities hold
even in the nonequilibrium relaxation process.
We can use finite-time and finite-size data in the scaling analysis.
The examples are Figs.~\ref{fig:chixi-ferro} and \ref{fig:chixi}.
Although the raw relaxation data 
(Figs.~\ref{fig:chixi-ferro-nama} and \ref{fig:chixi-nama})
exhibit  the finite-size effects,
we can scale them to one scaling line without the size effects.
The critical divergence is observed from very early stage of the 
nonequilibrium relaxation process,
if we scale the data by the dynamic correlation length.

The present dynamic-correlation-length-scaling analysis is regarded as
an extension of the finite-size-scaling analysis replacing the size $L$
with the dynamic correlation length $\xi(t)$.
We may consider that the finite-time relaxation data at a time $t$
corresponds to the equilibrium data of the size $L$ with $L=\xi(t)$.

Since the present scaling method is entirely based on the dynamic
correlation length, the definition is very important.
We found that a use of the definition based on the Ornstein-Zernike formula
is not appropriate in our scheme.
Since the formula is based on the mean-field approximation,
the present nonequilibrium process may be out of the applicable range
of the approximation.
The relaxation function of the dynamic correlation length
estimated by Eq.~(\ref{eq:1}) exhibits an extra increase before reaching
an equilibrium value, as shown in Figs.\ref{fig:isingxi} and \ref{fig:sgxi}.
The logarithmic slope, which is $1/z$, is then overestimated.
It affects a value of $\nu$ in the conventional finite-time-scaling analysis
of the nonequilibrium relaxation method.\cite{totareject}
The introduced definition, Eq.~(\ref{eq:2}), is based on a simple
exponential decay of the correlation function.
The relaxation function exhibits a normal behavior.
It algebraically increases with $t^{1/z}$, and eventually
converges to a finite-size value.
We also comment that the definition Eq.~(\ref{eq:2}) may be used even
in the equilibrium state where the mean-field approximation is not valid.

\acknowledgments
The author would like to thank
Professor Nobuyasu Ito and Professor Yasumasa Kanada 
for providing him with a fast random number generator RNDTIK.
This work is supported by Grant-in-Aid for Scientific Research from
the Ministry of Education, Culture, Sports, Science and Technology, Japan
 (No. 15540358).

\end{document}